\documentclass[twocolumn,showpacs,preprintnumbers,amsmath,amssymb,pre,aps]{revtex4-1}
\pdfoutput=1
\usepackage{graphicx,color}
\usepackage{url}
\usepackage[percent]{overpic}
\usepackage{rotating} 
\newlength{\defbaselineskip}
\newcommand{\setlinespacing}[1]%
           {\setlength{\baselineskip}{#1 \defbaselineskip}}

\begin{document}

\title{\normalsize Correcting for Bias of Molecular Confinement Parameters  Induced by Small-Time-Series Sample Sizes  in Single-Molecule Trajectories Containing  Measurement Noise}

\begin{abstract} 
Several single-molecule studies aim to reliably extract parameters  characterizing molecular confinement or transient kinetic trapping from experimental observations.   Pioneering works from single particle tracking (SPT) in membrane diffusion studies [Kusumi et al., \emph{Biophysical J.}, \textbf{65} (1993)] appealed to Mean
Square Displacement (MSD) tools for extracting diffusivity and other parameters quantifying the degree of confinement.  
More recently, the practical utility of  systematically treating
multiple noise sources (including noise induced by random photon counts) through  likelihood techniques have been more broadly realized in the SPT community.
However, bias induced by finite time series sample sizes (unavoidable in practice) has not received great attention.  
Mitigating parameter bias induced by finite sampling is important to any scientific endeavor aiming for high accuracy, but correcting 
for bias is also often an important
step in the construction of optimal parameter estimates. 
In this article, it is demonstrated how a popular model of confinement 
can be corrected for finite sample bias in situations where the underlying data exhibits 
Brownian diffusion and observations are measured with non-negligible experimental noise 
(e.g., noise induced by finite photon counts).  
The work of Tang and Chen [\emph{J. Econometrics}, \textbf{149} (2009)] 
is extended to correct for bias in the estimated  ``corral radius'' 
(a parameter commonly used to quantify confinement in SPT studies)
 in  the presence  of measurement noise.  
It is shown that the approach presented is capable of reliably extracting 
the corral radius using only hundreds of discretely sampled observations in situations where 
other methods (including MSD and Bayesian techniques) would encounter serious difficulties.  
The ability to accurately statistically characterize transient confinement suggests new techniques for 
quantifying confined and/or hop diffusion in complex environments.  
  \end{abstract}

\author{\small Christopher P. Calderon $^\dagger$}

\email{chris.calderon@numerica.us}
\affiliation{%
$^\dagger$ \small Numerica Corporation, 4850 Hahns Peak Drive, Loveland, Colorado, 80538 
}

\date{ \today}

\pacs{87.80.Nj, 87.10.Mn, 05.40Jc, 2.50.Tt, 5.45.Tp}

\maketitle

\section{Introduction} 

In many live cell applications, large scale cellular structures impose complex constraints on the motion of smaller biomolecules \cite{Schlessinger1977,Sako1994,Kusumi2005,Golding2006a,Destainville2006,Rohatgi2007,Saxton2007,Masson2009,Magdziarz2010,Nachury2010,Park2010,Weigel2011,Turkcan2012}.
Quantifying these effects from \emph{in vivo} observations is the goal of numerous experiments.
Fortunately, recent advances in microscopy and other single-molecule probes
 have substantially improved resolution in both time and space, so various complex kinetic constraints can be more quantitatively measured.

Fluorescence microscopy can be used to extract kinetic information from a sequence of point spread function (PSF) measurements \cite{Kim2006,Manley2008}.
  Pioneering efforts  \cite{Qian1991,Kusumi1993} aiming  to quantify transient ``corralling'' 
  parameters characterizing confinement induced by cytoskeletal and transmembrane protein structures \cite{Kusumi2005}  appealed to Mean Square Displacement (MSD) analyses.  
 Recently, the utility of  statistically motivated time series analysis have become more popular for analyzing single-molecule data.  These
 tools offer several advantages over traditional MSD-based  techniques.
For example, likelihood and Bayesian-based statistical analysis  methods permit one more flexibility in 
terms of inference decisions characterizing  noisy systems, and these schemes also provide more efficient estimation 
strategies \cite{vandervaart,Ober2004,Montiel2006,llglassy,Masson2009,Berglund2010,Michalet2012,Turkcan2012}.

Many of the first works utilizing likelihood-based analysis methods to analyze single particle tracking (SPT)  data ignored the effects of measurement noise (also referred to
as ``localization precision'' \cite{Thompson2010,Turkcan2012,Michalet2012}), but the importance of modeling this noise source has been
demonstrated in various works focused on analysis single-molecule data where measurement noise induced by the experimental apparatus is not negligible relative to
thermal fluctuations inherent to single-molecule measurements  \cite{SPAdsDNA,SPAfilter,Berglund2010,Michalet2012}. Ref. \cite{Michalet2012} provides a  discussion on issues associated with simultaneously quantifying measurement and molecular diffusion in SPT applications, but the focus of Ref. \cite{Michalet2012} is on optimal parameter estimation.
It is well-known that the maximum likelihood estimator is \emph{asymptotically} unbiased \cite{vandervaart} and achieves the Cramer-Rao lower bound  
when the assumed underlying model precisely matches the data generating mechanism producing observations \cite{Ober2004}.  In applications where tracking molecules for a long time is complicated due to crowding, photobleaching, and/or emitter ``blinking" \cite{Kim2006,Manley2008}, it is difficult to collect a large number of measurements (hence the asymptotic sampling regime is not encountered).    In PSF modeling, the appropriate parametric models have been more broadly agreed upon \cite{Ober2004}, but the ``correct'' stochastic model consistent with experimental single-molecule observations is a more delicate issue \cite{SPAgof,SPAfilter}. 
  Furthermore, even if observations are consistent with the assumed stochastic model, correcting for systematic bias  introduced by  finite sample sizes 
where observations contain both diffusive noise and measurement noise 
has not received great attention in the SPT literature.
Therefore estimators accurately quantifying finite sample bias (as opposed to asymptotically minimizing parameter variance or bias) are desirable when analyzing experimental trajectories.

This work introduces a bias correction scheme for extracting the ``corral radius'' \cite{Qian1991,Kusumi1993}.  This quantity is commonly  used to characterize confinement in biophysical applications  \cite{Saxton2007,Park2010,Turkcan2012}.  
Examples characteristic of sampling regimes encountered in fluorescence microscopy are presented, but the approach can be readily generalized to other time and length scales. The bias correction removes systematic errors induced by observing a short finite time series (enabling estimation in situations where an MSD curve is deemed too noisy) in contrast to removing artifacts of motion blur \cite{Destainville2006,Berglund2010}.  However the analysis presented explicitly shows how to map  estimated parameters to MSD curves, so previously proposed motion blur corrections for confinement \cite{Destainville2006,Berglund2010} can be used to   augment the  tools presented.

  Likelihood-based techniques \cite{hamilton,Shiriaev_limoexp_99,llglassy,Tang2009a,TSRV,SPAfilter, Ait-Sahalia2010} are employed throughout this article;
 such  methods  enable one to consider numerous  time series analysis 
tools  in physical and life science applications.  The author has found adopting statistically rigorous time series analysis tools from econometrics and computational finance helpful in statistically analyzing data from microscopic simulations \cite{SPA1, SPAJCTC, pudi} and single-molecule force manipulation experimental 
data where measurement noise is commensurate with thermal noise \cite{SPAfilter,SPAdsDNA,SPAfric}.  In this article, the   
relevance of recent likelihood-based 
tools \cite{Tang2009a,TSRV,Ait-Sahalia2010}  to SPT
 modeling is demonstrated.  Section \ref{sec:model} presents the stochastic differential equation (SDE) model considered, relates parameters extracted from these models to traditional MSD analyses, and introduces the basic tools utilized throughout.  The first figure and tables in Sec. \ref{sec:results} present the main results;  the remaining results explain and justify how a theory originally developed for estimating SDEs observed without measurement noise \cite{Tang2009a} can be modified and extended to handle the situation where measurement noise contaminates time series data.

\section{Methods} 
 \label{sec:model}
The underlying  position of a molecule will be denoted by $x$ and the noisy experimental observations will be denoted by $\psi$; the motion models considered take the following form:
\begin{align}
\label{eq:SDE}
dx_t=& -\nabla V(x_t) dt+\sigma dB_t &\\
\label{eq:meas}
\psi_{i}=& x_{i}+\epsilon_i ; \epsilon_i  \sim \mathcal{N}(0,R).
\end{align}
\noindent The above  is an SDE model \cite{risken} with a constant  diffusion coefficient  driven by a standard Brownian motion process $B_t$ (the subscripts denote a continuous time model) having a drift function determined by a potential $V(x)$.
The  measurements, $\psi_i$, in Eqn. \ref{eq:meas}  are contaminated by  noise, $\epsilon_i$, modeled as draws from  a Normal distribution  with mean zero and variance $R$ (denoted by $\mathcal{N}(0,R)$). In SPT applications, the effective measurement noise (i.e., localization precision) is often quantified by $R^{1/2}$. The measurement noise is typically assumed to be an independent and identically distributed (i.i.d.) random number sequence,  and the variance $R$ is assumed unknown \emph{a priori} (the model also assumes statistical independence of $x$ and $\epsilon$).
  The integer subscript $i$ denotes that trajectory observations are made at  discrete times and 
$t_{i+1}-t_i = \Delta t \ \forall \ i$.  Since typical SPT calculations assume independence between spatial coordinates \cite{Kusumi1993,Destainville2006,Park2010,Michalet2012}, we will restrict attention to analyzing the 1D version of Eqn. \ref{eq:SDE}; hence the diffusion coefficient is $D:=\frac{\sigma^2}{2}$.

For  $V(x)$, two different functional forms will be considered: (i) $V(x)=0$ for 
$|x|<L/2$ and $V(x)=\infty$ for $|x|\ge L/2$ which we refer to as reflected Brownian motion (RBM); in this case the parameters needed to completely characterize particle motion are $(L,\sigma,R)$ and (ii) $V(x)=\frac{1}{2}\kappa x^2$ which we label as the Ornstein-Uhlenbeck (OU) process (also known as the Vasicek model \cite{Tang2009a}); parameters requiring estimation in this case are $(\kappa,\sigma,R)$. 
Both potentials mentioned above have been considered in confined membrane diffusion studies \cite{Kusumi1993,Destainville2006,Turkcan2012}.  Kusumi \emph{et al.} demonstrated how the MSD asymptotically approaches $\frac{L^2}{6}$
in the RBM model;  extraction of $L$ from data  is still a common technique for quantifying confinement in SPT studies \cite{Destainville2006,Park2010,Turkcan2012}  (the 1D corral radius is defined by $\sqrt{\frac{L^2}{6}}$).  
The MSD corresponding to an ergodic OU process observed with infinite time  for $\delta$ time units between adjacent observations can (see Appendix) be shown  to be
$\frac{\sigma^2}{\kappa} \big(1- e^{-\kappa \delta}\big)$.

If the two models under consideration have identical diffusion coefficients, then setting 
 $\kappa=\frac{6\sigma^2}{L^2}$ is one way to match asymptotic MSD parameters;  in the confined regime, this relation  also allows one to map $\kappa$ of the OU model onto the corresponding $L$ parameter in the RBM model.  The Appendix displays representative trajectories and also compares the entire MSD for OU and RBM models driven by the same Brownian noise realizations. In the measurement noise free case ($R=0$) with large samples, the RBM and OU processes are easy to qualitatively and quantitatively distinguish. When  measurement noise is present ($R>0$), the two scenarios are much harder to  distinguish if one only has access to a few hundred observations of each trajectory.  
 Using only 100-400 observations, hypothesis testing tools \cite{hong} cannot statistically distinguish the two models in parameter regimes of relevance to many SPT studies.  
 The time series length required to obtain adequate power to statistically distinguish RBM from the OU processes observed with measurement noise is larger than typical track lengths encountered in practice.
 If statistical signature of other more complex noise cannot be systematically detected in the sample sizes commonly encountered in practice \cite{Golding2006a,Saxton2007,klafter08,Magdziarz2010,Weigel2011}, one should consider modeling with the OU process because of statistical advantages this  process offers when analyzing experimental data (these are discussed in the next subsection).  The advantages (from a physical standpoint) of applying detailed time series analysis to  short trajectories experiencing transient confinement are  discussed in Sec. \ref{sec:conclusions}.  \\

\textbf{Advantages Afforded by the OU Model}

 The discrete time analog of Eqn. \ref{eq:SDE} for the OU model is:
\begin{align}
\label{eq:SDEd}
x_i=&   F x_{i-1}+\eta_{i-1}  \ ;& \eta_{i-1} \sim \mathcal{N}(0,Q) \hfill \\
\label{eq:measd}
\psi_{i}=& x_{i}+\epsilon_i  \  ;& \epsilon_i \sim \mathcal{N}(0,R) \hfill,
\end{align}

\noindent where $F\equiv e^{-\kappa\Delta t}$ and $Q \equiv \frac{\sigma^2}{2\kappa}( 1- e^{-2\kappa\Delta t})$ \cite{Tang2009a}.  This relation  allows one to readily use the Kalman filter estimation framework \cite{hamilton}.  
Maximum likelihood estimation (MLE) of the parameters  completely characterizing the stationary OU process can be computed from the observable measurements $\{ \psi_i\}_{i=0}^N$   \cite{hamilton,SPAdsDNA,SPAfric}.  This permits efficient  estimation in situations where sample sizes for an MSD analysis are difficult to reliably extract  and statistically characterize (see Appendix Fig. \ref{fig:A2}).  The Gaussian structure of the OU process also enables one to exploit a variety of other powerful tools that can be used to analyze this type of stochastic process \cite{hamilton}, including goodness-of-fit testing (checking model assumptions against data directly \cite{SPAdsDNA,SPAfilter,SPAgof}),  exact rate of convergence analysis under stationary and non-stationary sampling \cite{Shiriaev_limoexp_99}, and bias correction.
For example, Tang and Chen \cite{Tang2009a} demonstrate how to remove bias from MLEs computed using finite sample sizes in the case where the $x_i$'s are directly observed (i.,e., $R=0$).  In the stationary case ($\kappa>0$), it can be shown using moment bounds for weakly dependent sequences 
\cite{yokoyama80,billingsley,Tang2009a} that: 

\begin{align}
\label{eq:bias}
\mathbb{E}[{  \hat{\kappa}}]= &{ {\kappa}} \ + \\
\nonumber & \frac{1}{N\Delta t} \big(   \frac{5}{2}+ e^{\kappa \Delta t}+ \frac{1}{2}e^{2\kappa\Delta t} \big) + \mathcal{O}(\frac{1}{N^2}),
\end{align}

\noindent where $\mathbb{E}[{  \hat{\kappa}}]$ denotes the expectation of the MLE of $\kappa$ (the MLE is denoted by $\hat{\kappa}$). The other terms quantify the expected bias induced by finite $N$. 
  $\hat{\kappa}$ is often the dominant source of bias
when the relation  $L=\sqrt{\frac{6\sigma^2}{\kappa}}$ is used to extract the corral radius from  OU parameter estimates
in the sampling regimes studied (e.g., results obtained by plugging in the corrections to $\sigma^2$ reported in Ref. \cite{Tang2009a} did not affect results). 

Before moving onto the case where $R>0$, it is worth reviewing a  classic first order autoregressive time series model \cite{hamilton,Shiriaev_limoexp_99} where $x_i= F x_{i-1}+\eta_{i-1}$ where $\eta_{i-1} \sim \mathcal{N}(0,Q)$; the interest is in estimating $F$ ($Q$ is considered frozen and to be nuisance parameter).
For notational simplicity set $Q=1$ and $x_0=0$. 
 In this case, for given  sample of size $N$ (also referred to as the ``track length'') the standard likelihood equation is:
\begin{align}
p_F(x_1,x_2,\ldots,x_N)=(2\pi)^{-\frac{N}{2}}\exp\big( -\frac{1}{2}\sum\limits_{i=1}^{N} (x_{i}-Fx_{i-1})^2 \big)
\end{align}
Taking the logarithm, expanding the quadratic terms, and setting the derivative of the expression above with respect to $F$ equal to zero provides the following estimator \cite{Shiriaev_limoexp_99}:
\begin{align}
\label{eq:Fmle}
\hat{F}=\frac{\sum\limits_{i=1}^{N} x_{i}x_{i-1}}{\sum\limits_{i=1}^{N}x_{i-1}^2}
\end{align}
 In the presence of measurement noise, the above suggests a naive suboptimal (denoted by a tilde) estimator:
 \begin{align}
 \label{eq:qFmle}
 \tilde{F}=\frac{\sum\limits_{i=1}^{N} (x_{i}+\epsilon_i)(x_{i-1}+\epsilon_{i-1})}{\big(\sum\limits_{i=1}^{N}(x_{i-1}+\epsilon_{i-1})^2\big)-N\tilde{R}}
 \end{align}
\noindent where $\tilde{R}$ is an independent estimate of the measurement noise variance. Recall that the measurement noise is assumed i.i.d., so if $\tilde{R}$ is asymptotically consistent and $Q$ is fixed, $\tilde{F}$ is asymptotically consistent since the cross-term sums involving $\epsilon$ and $x$ tend to zero and become  insignificant relative to the other non-zero sums in the $\kappa>0$ case under study.  The problem with this approach is that the estimator is suboptimal (the cross-terms increase estimation variance).  Unfortunately, the estimator above also requires one to construct a consistent $\tilde{R}$ (this can alternatively come from a prior, but this will likely introduce bias which is hard to quantify).  Furthermore, if one uses estimators ignoring confinement effects, new systematic biases (on top of inherent finite sample bias associated with estimating $\kappa$) can be introduced.  This phenomenon is demonstrated by example in the Results.

 In the Kalman filter framework considered, the innovation likelihood (Appendix Eqn. \ref{eq:innov}) has an approximate autoregressive \cite{hamilton} form if the filter covariance reaches steady state quickly.  If a stationary OU process is deemed adequate to describe experimental observations and the Kalman filter covariance sequences reaches its steady state value rapidly, 
 then analysis in Ref. \cite{Tang2009a} can be applied to study the expected finite sample bias  of $\hat{\kappa}$. 
 When one jointly estimates the MLE parameters associated with Eqn. \ref{eq:SDE} by optimizing the innovation likelihood, 
 one effectively returns to the situation in Eqn. \ref{eq:Fmle} where the estimates of $F$   can be extracted without knowledge of the value of the constant noise parameters.   In the Results (Fig. \ref{fig:S}), the convergence of matrices characterizing the Kalman filter are demonstrated.  

In what follows, it is shown how   plugging the MLE's  (obtained by maximizing Eqn. \ref{eq:innov}) into Eqn. \ref{eq:bias}  can significantly reduce  bias from
parameter estimates obtained with small $N$ in situations of relevance to SPT tracking (the approach avoids specifying the ``lag parameter'' plaguing MSD-based analyses \cite{Michalet2012}).
The approach  is demonstrated to accurately infer both $\kappa$ and the effective $L$ (corral radius) if data is generated using either the OU model (correct model specification) or the RBM (model misspecification).

\section{Results}
\label{sec:results}

\begin{figure}[htb]
\center
\centering
\begin{minipage}[b]{.99\linewidth}
\def\pw{.85}
\begin{overpic}[width=\pw\textwidth]{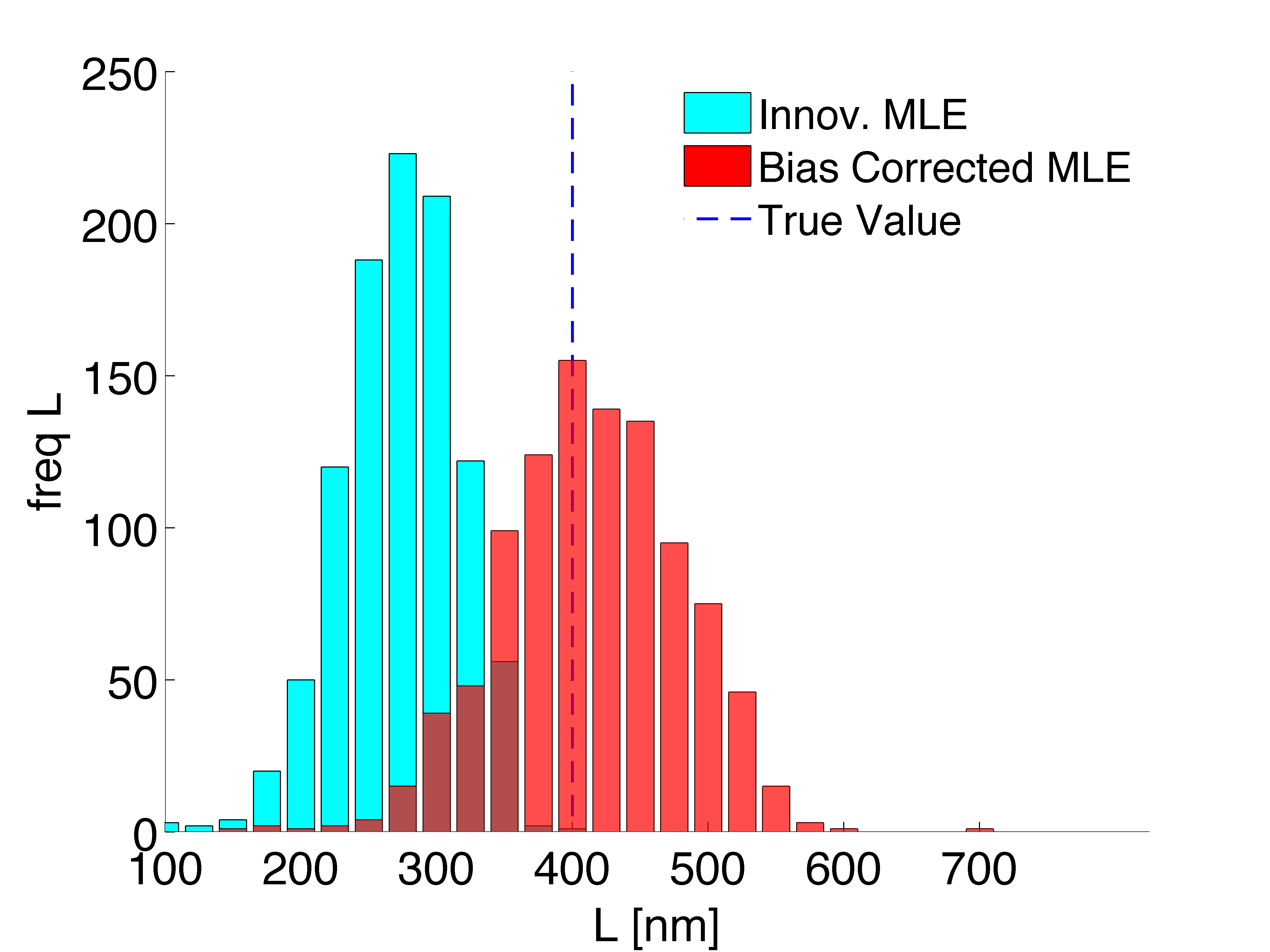}
\put(15,60){\Large \color{black}(a)}
\end{overpic}
\begin{overpic}[width=\pw\textwidth]{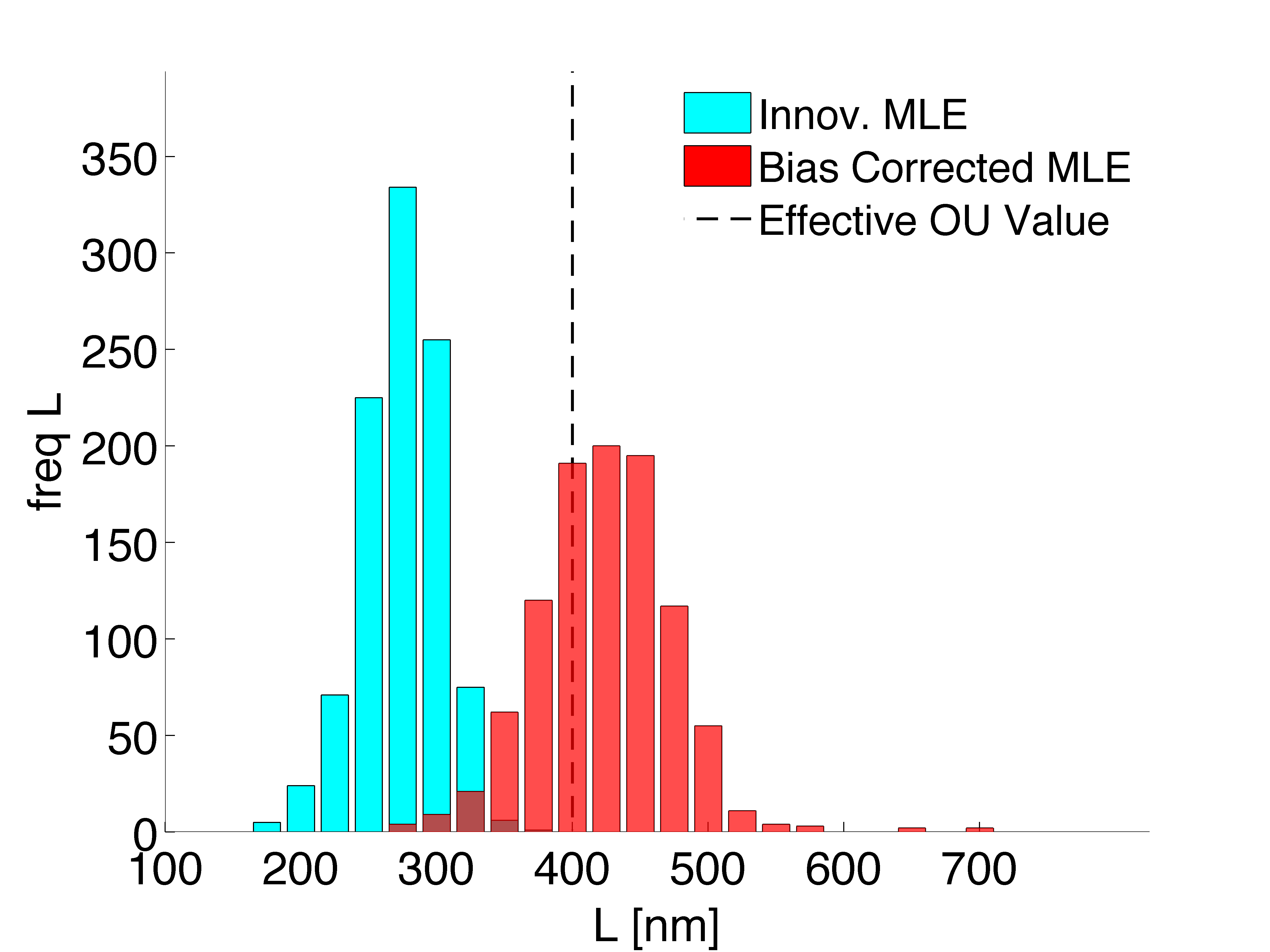}
\put(15,60){\Large \color{black}(b)}
\end{overpic}

  \end{minipage}
\centerline{\footnotesize }
\caption{
\footnotesize 
(Color online) Raw MLE and bias corrected corral radius estimate, $\hat{L}$,  obtained by extracting the parameters from time series of length 100 sampled every $25 ms$ (this was repeated for 1000 Monte Carlo trials; the histogram displays 1000 $\hat{L}$'s). In (a), the OU process (with known parameters) generates observations.  In (b), the same Brownian motion paths used to generate OU trajectories are used to construct RBM paths.  In both cases, a single measurement noise random number stream was added to each trajectory. The use of the same underlying Brownian path and measurement noise sequence was used to reduce random variation and facilitate quantifying systematic errors. 
 }
\label{fig:1}
\end{figure}

Figure \ref{fig:1} presents a histogram of the raw  estimate of the corral radius obtained via the relation 
$\hat{L}=\sqrt{    \frac{6\hat{\sigma}^2}{\hat{\kappa}}     }$ for the case where 1000 Monte Carlo simulations with 
$L=400 nm, D=0.2\mu m^2/2, R^{1/2}=50 nm$, $\Delta t = 25 ms$ and $N=100$ observations of $\psi$ were used to generate data. From this data, the   
 parameter estimates  characterizing the model in Eqn. \ref{eq:SDE} were extracted.  Corral radius parameters are inferred using the MLE and the bias corrected parameter estimates for two different data generating processes. In the top panel, the OU process generates data; in the bottom panel, the RBM process generates data (here there is model mismatch).  The bias induced by only observing 100 time series is effectively removed in both cases.  Appendix Fig. \ref{fig:A2} displays representative  trajectories of $x$, $\psi$, and the empirical MSD associated with these trajectories.

\begin{table} [htb]
\center
\caption{\label{tab:T1}  Corral radius estimates with innovation MLE and bias corrected MLE (below referred to as ``Classic Innov.'',``Bias Cor.'', respectively).  The columns labeled with $\hat{L}$
contain the average parameter estimate obtained by analyzing 200 Monte Carlo trajectories each containing 400 observations spaced by $\Delta t =25 \ ms$ (the number in parenthesis reports standard deviation).  The column labeled error reports the mean minus known true corral radius.  In this table $D=0.2 \mu m^2/s$ and $R^{1/2}=25 nm$. \baselineskip=10pt  }
 \center
\begin{tabular}{|l|*{4}{c|}} \hline

& \multicolumn{2}{c|}{ OU } & \multicolumn{2}{c|}{ RBM}  \\ \hline \hline
Estimator & $\hat{L} \ [nm]$  & Error  & $\hat{L} \ [nm] $ & Error \\ \hline 
& \multicolumn{4}{c|}{ $L=250 nm$}   \\ \hline \hline
Classic Innov. & 169.31 (16.07) & -80.69  & 168.19 (12.56) & -81.81  \\ \hline

{ Bias Cor.}  & {241.40 (22.54 )} & {-8.60} & {240.24 (17.56 )} & {-9.76} \\ \hline

& \multicolumn{4}{c|}{ $L=400 nm$}   \\ \hline \hline
Classic Innov.  & 277.09 (18.38) & -122.91 & 275.69 (10.63) & -124.31  \\ \hline

{ Bias Cor.} & {399.09 (26.89 )} & {-0.91 } & {398.21 (15.39 )} & {-1.79} \\ \hline

& \multicolumn{4}{c|}{ $L=500 nm$}   \\ \hline \hline
Classic Innov.  & 344.25 (25.75) & -155.75 & 343.70 (14.83) & -156.30 \\ \hline

{ Bias Cor.} & {500.35 (38.69 )} & { 0.35 } & {501.73 (22.54 )} & { 1.73 } \\ \hline

\end{tabular}
\end{table}

\begin{table} [htb]
\center
\caption{\label{tab:T2}  Same as Table \ref{tab:T1} except  $D=0.02 \mu m^2/s$. \baselineskip=10pt  }
 \center
\begin{tabular}{|l|*{4}{c|}} \hline

& \multicolumn{2}{c|}{ OU } & \multicolumn{2}{c|}{ RBM}  \\ \hline \hline
Estimator & $\hat{L} \ [nm]$  & Error  & $\hat{L} \ [nm] $ & Error \\ \hline 
& \multicolumn{4}{c|}{ $L=250 nm$}   \\ \hline \hline
Classic Innov. & 169.43 (19.17) & -80.57  & 170.70 (12.24) & -79.30  \\ \hline

{ Bias Cor.}  & {253.54 (31.84)} & { 3.54} & {258.49 (21.06)} & { 8.49} \\ \hline

& \multicolumn{4}{c|}{ $L=400 nm$}   \\ \hline \hline
Classic Innov.  & 253.84 (46.88) & -146.16  & 257.78 (34.79) & -142.22  \\ \hline

{ Bias Cor.} & {418.74 (118.61)} & {18.74 } & {437.48 (90.10)} & {37.48 } \\ \hline

& \multicolumn{4}{c|}{ $L=500 nm$}   \\ \hline \hline
Classic Innov.  & 310.85 (61.59) & -189.15  & 308.72 (57.53) & -191.28  \\ \hline

{ Bias Cor.} & {582.40 (220.27)} & {82.40} & {603.12 (244.42)} & {103.12} \\ \hline

\end{tabular}
\end{table}

Tables \ref{tab:T1}-\ref{tab:T2} present similar results, but vary the system and sampling parameters.  The parameters explored were motivated by SPT studies.  Even for $N=400$, substantial bias exists in the asymptotically efficient MLE.  Bias reduction comes at the cost of variation as can be observed by the reported standard deviations.  However, using the OU model structure allows one to use a wealth of quantitative tools for understanding experimental data analysis.  The main results have now been presented, what follows expands on technical details and on the domain of applicability of the bias removal approach.

\begin{figure}[htb]
\center
\centering
\begin{minipage}[b]{.99\linewidth}
\def\pw{.85}
  \begin{overpic}[width=\pw\textwidth]{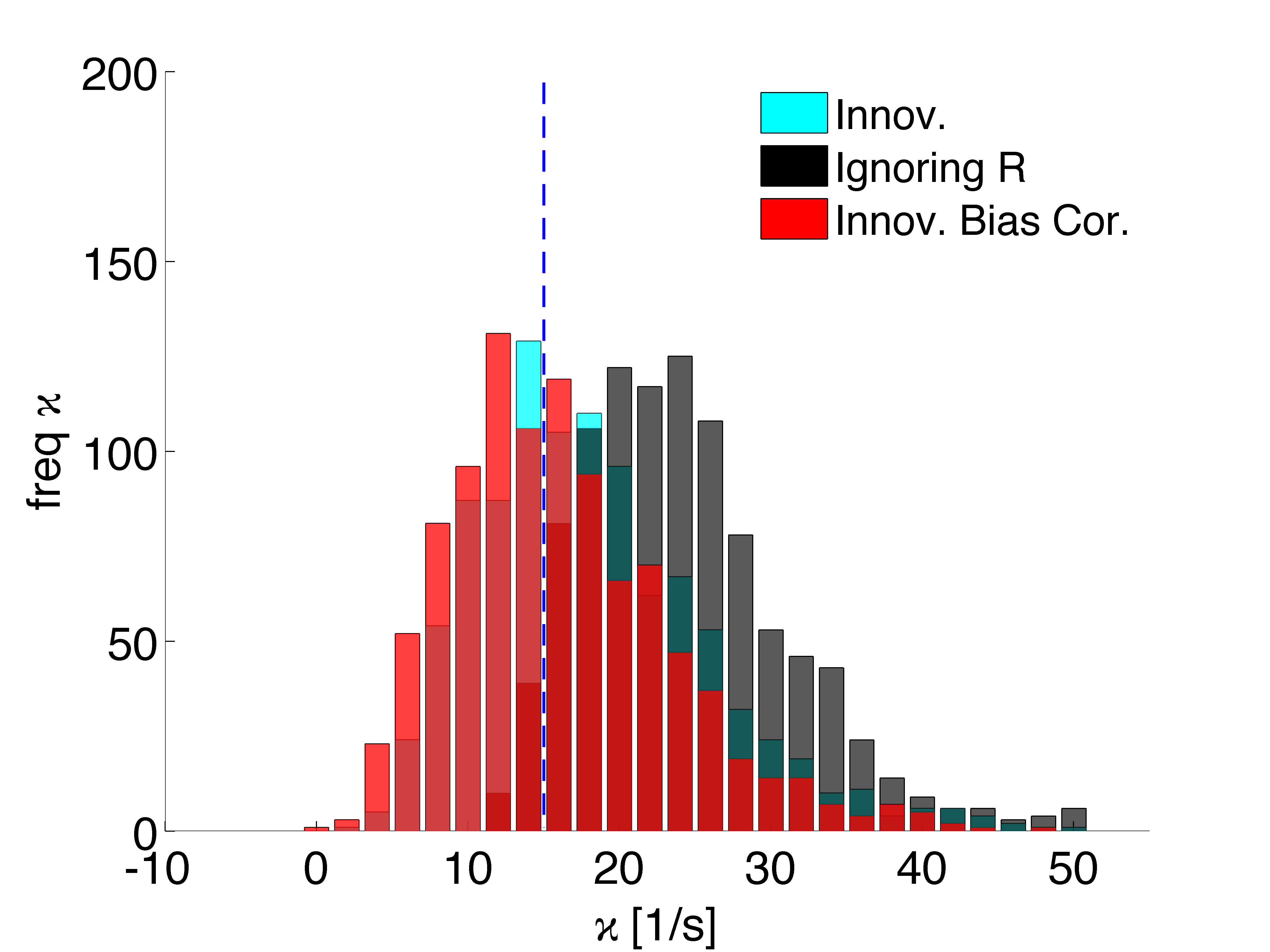}
\put(15,60){\Large \color{black}(a)}
\end{overpic}
\begin{overpic}[width=\pw\textwidth]{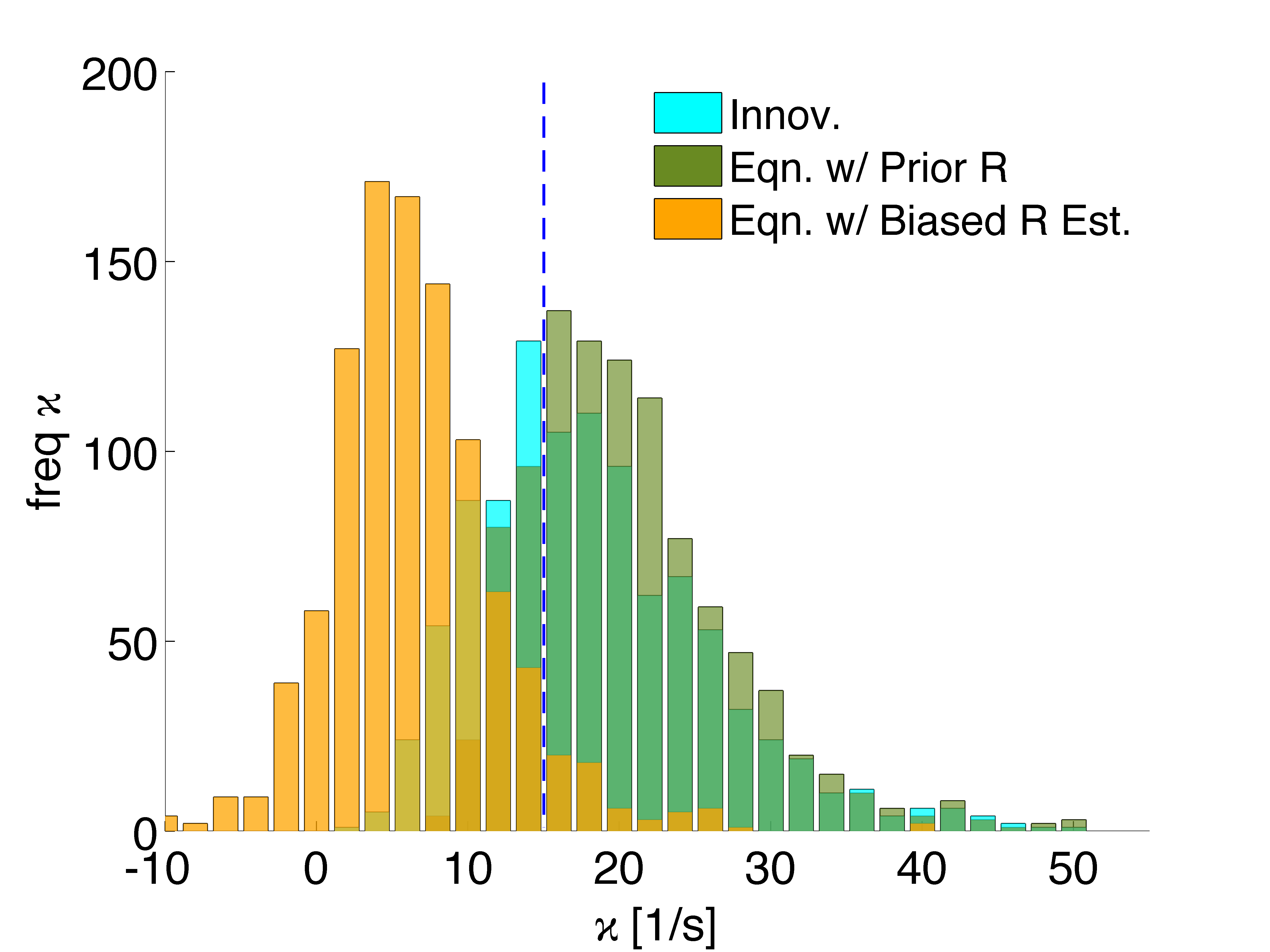}
\put(15,60){\Large \color{black}(b)}
\end{overpic}
  \end{minipage}
\centerline{\footnotesize }
\caption{
\footnotesize 
(Color online) Distribution of $\hat{\kappa}$ obtained when the  OU process with measurement noise generated observations (results correspond to $L$ histogram shown in Fig. \ref{fig:1}). Fixed true value of $\kappa$ denoted by vertical dashed line.  Two different classes of estimators were used: (a) MLE-based estimators that use a likelihood with the correct model and (b) ``Other'' estimators use an assumed prior input for $\tilde{R}$ (the true $R^{1/2}$ is 50 nm, but $\tilde{R}^{1/2}$ is set to 40 nm since knowledge of the precise effective measurement noise is difficult to accurately quantify \cite{Berglund2010} in SPT applications) and a suboptimal estimator in Eqn. \ref{eq:qFmle} (for $\tilde{R}$ we plug-in the noise estimate obtained using code associated with Ref. \cite{Michalet2012}).  In both cases, the Innovation MLE without bias correction (labeled as ``Innov.'') serves as the reference histogram.
 }
\label{fig:2}
\end{figure}

Figure \ref{fig:2} presents the distribution of the estimated $\kappa$ for a variety of estimators.  
The focus is on $\kappa$ since this is often the major source of variation in $L$ estimated from short time series \cite{Tang2009a}.
The top panel displays three estimators; (i) the raw MLE associated with the OU process where $R$ is assumed zero \cite{Tang2009a}; (ii) the MLE obtained by jointly extracting the $\kappa,\sigma^2,$ and $R$ that minimize the innovation likelihood \cite{SPAfric} (see Eqn. \ref{eq:innov}) and; (iii) using the bias correction of Tang and Chen \cite{Tang2009a} applied to the output of (ii).   The average of the $\hat{\kappa}$ distributions for the three cases are 24.3, 18.2, and 16.0 $\frac{1}{s}$, respectively (the true value is 15 $\frac{1}{s}$).  The difference may seem small, but recall that the estimated corral radius depends nonlinearly 
on $\hat{\kappa}$ (hence the amplified difference in $\hat{L}$).

The bottom panel in Fig. \ref{fig:2} uses ``other'' suboptimal estimators of $\kappa$.  In one case $R$ is  assumed to be known accurately \emph{a priori};  here $\tilde{R}^{1/2}=0.8R^{1/2}=40nm$ was  used along with Eqn. \ref{eq:qFmle} to estimate $\kappa$.  Since accurate \emph{a priori} knowledge of $R$ can be a questionable assumption in SPT studies, 
we also show results of applying Eqn. \ref{eq:qFmle} in conjunction with  the estimator reported in Ref. \cite{Berglund2010} to extract $\tilde{R}$ from the data;  here $R$ is biased because the estimator in Ref. \cite{Berglund2010} was designed for the case where no forces or confinement constraints affect particle dynamics (the average  of the estimates of $R$ assuming the model in Ref. \cite{Berglund2010} was 72.4 $nm$ for this data set; note that Refs. \cite{Berglund2010,Michalet2012} warn that the estimator is not valid if constraint forces are present).  

The average  MLE (without bias correction) for the diffusion and measurement noise was (0.23 $\mu m^2/s$, 41.9 $nm$), that using the estimator from Ref. \cite{Berglund2010} was (0.07 $\mu m^2/s$, 71.3 $nm$), and the true value for the OU data generating process was (0.20 $\mu m^2/s$, 50.0 $nm$).
 Note that the arguments appealed to in this paper to explain the validity of the bias correction of Ref. \cite{Tang2009a} in conjunction with the Kalman filter's innovation likelihood \cite{hamilton,SPAfilter}  (relevant expressions shown in Eqn. \ref{eq:innov})
are not directly applicable to the  bias correction of the other parameters  reported in \cite{Tang2009a}  when $R>0$. 
Analysis of the bias and variance of parameters $\sigma$ and $R$ are more involved due  to iterations introduced by the Kalman filter's update and forecast steps; this analysis is beyond the scope of this work.

Application of various estimators of OU parameters to data generated by both the OU and RBM (a misspecified model) processes, 
was carried out for two reasons: (i) to emphasize that certain estimators can induce subtle systematic biases and (ii) to stress that likelihood-based inference permits other analysis tools beyond estimation.   Bias correction is possible in addition to other techniques.  For example, detecting confinement from observations  using visual inspection of the short trajectories is problematic (Fig. \ref{fig:A2} shows how even in the $R=0$ case, distinguishing RBM from the OU process is difficult with $N=100$).  However, goodness-of-fit testing can be employed \cite{hong,SPAgof}.  Applying the technique of Hong and Li \cite{hong} (more specifically computing the $M(1,1)$ test statistic) allows one to reject $\approx 20\%$ of the trajectories assuming the so-called directed diffusion  model (i.e., constant diffusion, measurement noise and velocity \cite{Park2010,Thompson2010}, but $\kappa=0$) even with $N=100$.  There is overwhelming statistical evidence for larger $N$ cases (the average $p-$value obtained assuming the directed diffusion plus measurement noise model was $<5\times10^{-4}$ for all $N=400$ cases considered).  This demonstrates that the test has power to detect kinetic signatures of confinement in the presence of diffusive plus measurement noise in regimes of interest to SPT studies (if the model was not rejected one can entertain using models involving fewer parameters e.g., see Refs. \cite{Berglund2010,Michalet2012}).
 The case where we assumed an OU model, but an RBM model actually generated the data (model misspecification), was statistically indistinguishable using tests in Ref. \cite{hong} from the case where the OU model generated data. 
Since there is no evidence in the raw observational data favoring one model over the other, and both models produce similar estimates of the quantity of interest (the corral radius), 
it is attractive to use the OU modeling viewpoint since a substantial body of literature exists for analyzing data generated by this type of stochastic process \cite{hamilton,Shiriaev_limoexp_99,llglassy,Tang2009a,TSRV,SPAfilter, Ait-Sahalia2010}.

\begin{figure}[htb]
\center
\centering
\def\pw{.85}
  \includegraphics[width=\pw\linewidth]{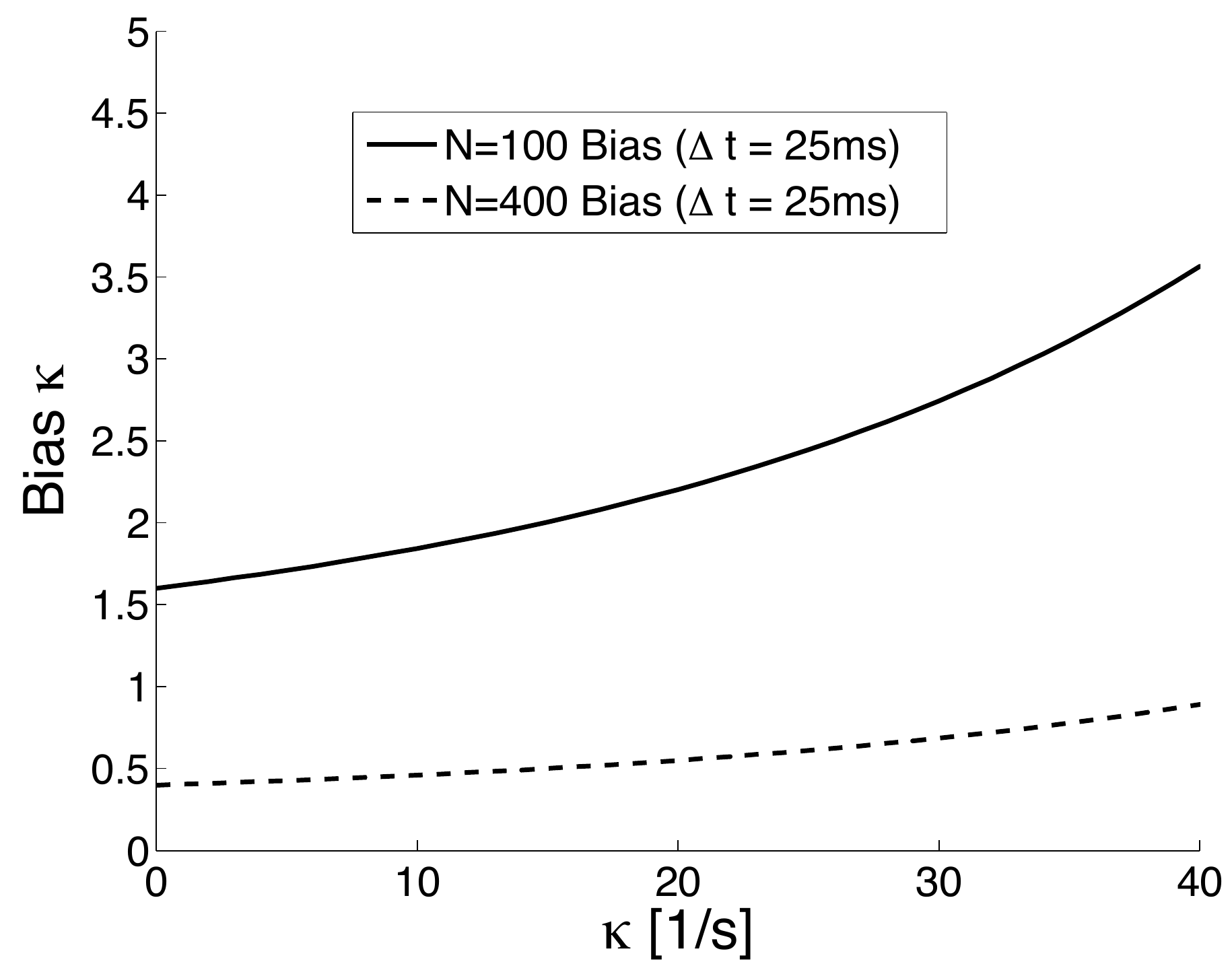}
\caption{
\footnotesize  (Color online)
Bias vs. $\kappa$ for two different sample sizes $N$ using Eqn. \ref{eq:bias} ($\Delta t = 25 ms$ and measurement noise magnitude $25 nm$).
 }
\label{fig:3}
\end{figure}

Figure \ref{fig:3} provides another example of analysis tools that are made available from likelihood-based analyses.  Here $\kappa$ is plotted against the truncated bias expansions in Eqn. \ref{eq:bias} (taken from Tang and Chen \cite{Tang2009a}) for a fixed $\Delta t$ and two sample sizes $N$.  Note how as $\kappa$ decreases, the fraction of bias increases rapidly.  Also note, that as $\kappa$ decreases, there is a higher likelihood of an MLE parameter estimate being near or less than zero (even for a truly stationary process where the underlying data has $\kappa>0$).  A high value of $\kappa$ suggest weak ``corralling'' since $L$ is inversely related to $\kappa$. The inverse dependence also causes the inflated standard deviation for $L=500 nm$ since a small fraction of estimated $\hat{\kappa}$ are near zero (also note that the median $\hat{L}$'s corresponding to the $L=500nm$ row of Tab. \ref{tab:T2} were 540.3 and 582.4; this suggest that these estimates in the tail of the estimated parameter distribution substantially influenced the observed mean). 
 Although one can remove expected bias, if the fraction of bias is large relative to the signal then other factors can complicate bias correction.  For example,  (i) higher order terms in the expected bias expansion can become more important; (ii) inherent parameter uncertainty in the point estimate substantially affects the expected bias. 
Therefore, plots like Fig. \ref{fig:3} allow researchers to  quantitatively determine when other factors influencing
the bias correction scheme need to be considered.

 \textbf{Conditions Required for Bias Removal}

\begin{figure}[htb]
\center
\centering
\def\pw{.825}
  \includegraphics[width=\pw\linewidth]{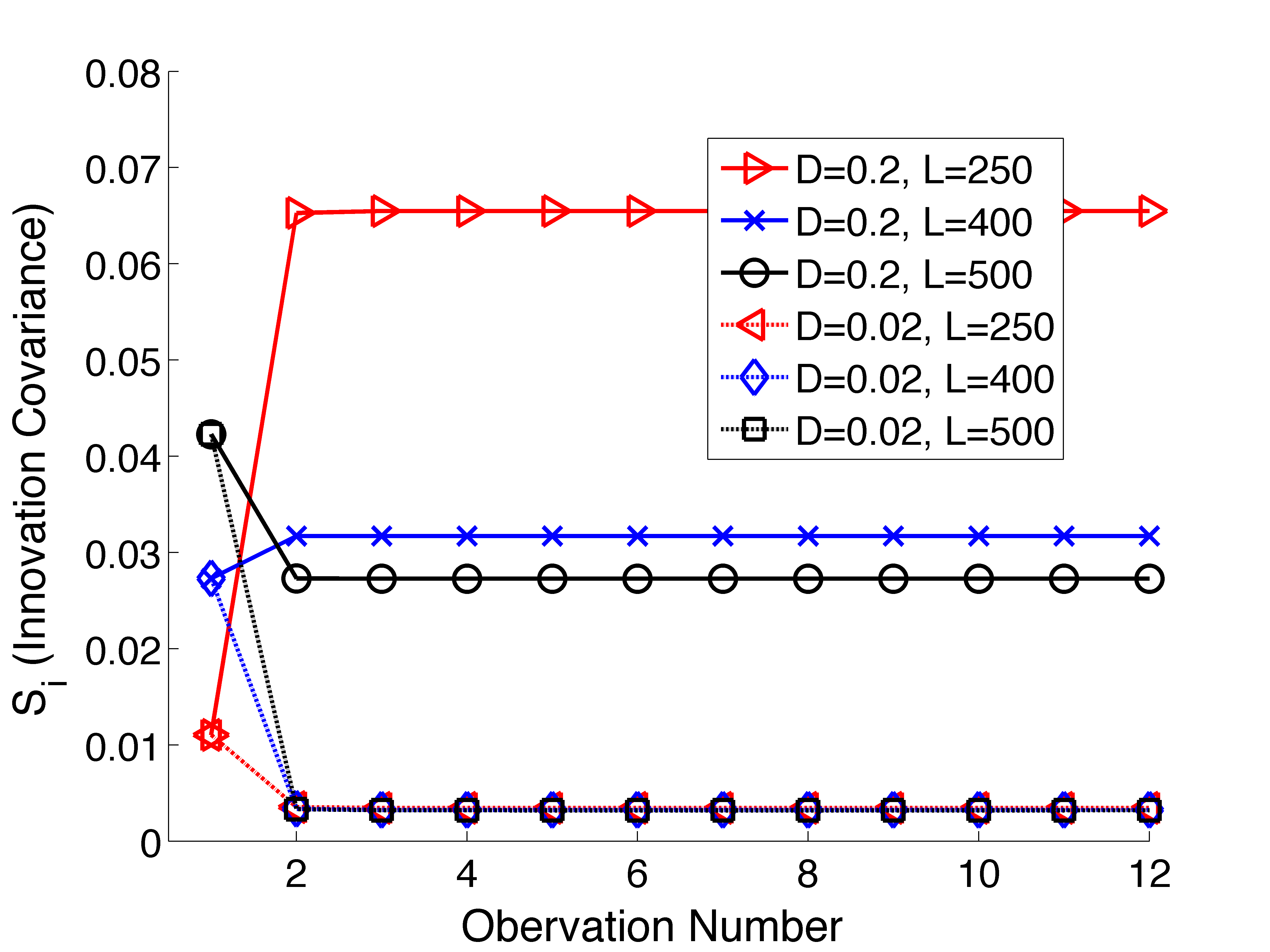}
\caption{
\footnotesize (Color online) 
Innovation covariance (see Eqn. \ref{eq:innov}) convergence rates.  Plots were obtained by plugging in exact data generating parameters studied in Tables \ref{tab:T1}-\ref{tab:T2}.  The plot illustrates that the filter quickly reaches steady state relative to the sample size $N$.
 }
\label{fig:S}
\end{figure}

The Kalman filter's constant noise assumption 
(i.e., the covariance of the innovation sequence, $S_i$, in Eqn. \ref{eq:innov}) 
needs to be tested in order for the analysis of Tang and Chen \cite{Tang2009a}
to be accurate for the expected bias in $\hat{\kappa}$.  Figure \ref{fig:S} illustrates that this is indeed the case for the parameter regimes under study.
Note that we intentionally ignored the estimation of the mean of the OU process (the mean was set to zero), this simplifies analyzing
the effect of $\kappa$ on the autoregressive parameter $(F=e^{-\kappa \Delta t})$ under the assumption of a constant  innovation covariance.
  The mean zero OU process does not restrict utility (with extra effort one can analyze the joint mean and $\kappa$ estimates and in practice one can simply demean the $\psi$ series using the empirically average of $\{\psi_i\}_{i=0}^N$).

Beyond testing constancy of the covariance of the innovation, one should verify that the bias in the estimated $R$ and $\sigma^2$ is small in relation to the bias in the estimated $\kappa$.  This can be achieved via simulation if need be. If bias in $R$ and $\sigma^2$ is determined to be significant, new analytical or numerical bias removal schemes should be considered. 
Bootstrap techniques can  be leveraged to quantify variance 
 and bias
in more complex SDE models \cite{davison1997bootstrap} (e.g., bootstrap techniques can check if the sample size is deemed too small for using a particular estimator).

  When $\Delta t$ is decreased with $R$ fixed, measurement noise often becomes a more dominant part of the single-molecule signal and the  bias in $R$ is relatively small \cite{TSRV,SPAdsDNA}. 
  In the OU model, the influence of $\kappa$  on $Q$ is $\mathcal{O}(\Delta t^2)$   since a Taylor expansion in $\Delta t$ shows $Q=\frac{\sigma^2}{2\kappa}\big(2\kappa \Delta t  + \mathcal{O}(\Delta t^2)\big)$.
    In the  small $\Delta t$ limit, one can leverage existing nonparametric tools for estimation and inference \cite{Ait-Sahalia2010}.  However, the parameter regime explored here is one in which $\kappa$'s influence is not small relative to $\Delta t$ (otherwise, the approach in Ref. \cite{Michalet2012} would predict more accurate $R$ estimates).  In this study, it was empirically demonstrated that the bias connected to the innovation MLE of $R$ is small  relative
    to that of $\kappa$ in several parameter regimes of relevance to SPT modeling (no bias correction was applied to $\hat{R}$).

Note that the bias correction scheme presented also depends heavily on the stationarity assumption of both the state
and on the innovation sequence.
If stationarity of $x$ is questionable, computing a corral radius should be reconsidered.  
More formally,  unit root tests \cite{hamilton} (adjusted to account for measurement noise) can be used
to check the Brownian motion vs. stationary OU models.
To more generally test stationarity of the mean or covariance of the observed measurements, other testing procedures can also be considered, e.g. \cite{Koutris2008}.   If the (bias corrected) estimate  along with the associated parameter uncertainty suggest $\hat{\kappa}$ is near zero, the suitability of a confined diffusion model needs to be carefully reevaluated. 

Finally, if all conditions mentioned in this subsection are met \emph{and} the Kalman filter corresponding to the OU process is an adequate model of the observations (an assumption tested here with time series hypothesis testing \cite{hong}), then
  the state and innovation noise residuals have mean zero (these residuals make up stationary process under the conditions above). 
  The filter and measurement noise sources make the innovation sequence different than classic order one autoregressive process, 
  but the additional noise terms
   do not substantially influence the first order expansion of the expected bias of $\hat{\kappa}$.  The effects of the additional
   noise terms 
   are lowest when the scalar gain, 
  $K_i$ (see Eqn. \ref{eq:innov}), is close to one (a standard autoregressive process generates the data when $K_i=1$). 
  In small time series sample sizes, 
  even when $K_i < \frac{1}{2}$, the bias correction can be shown to be accurate since the effects of parameter
  uncertainty  tend to dominate the additional noise associated with the filtered state estimates.  Furthermore, the MLE
  parameters are found by jointly optimizing the Eqn. \ref{eq:innov} given data, 
  but when the innovation covariance quickly reaches steady state, 
  the analysis of Tang and Chen \cite{Tang2009a} is relevant to understanding  the expected bias in $\hat{\kappa}$.

\section{Conclusions}
\label{sec:conclusions}
Simulations were used to demonstrate that kinetic confinement parameters could be accurately extracted 
from  
relatively short   time series ($100 \le N \le 400$)
containing both inherent diffusive and measurement noise (the latter prevents direct observation 
of position).   A bias correction scheme expanding off of Ref. \cite{Tang2009a} was presented;  
it was demonstrated that the scheme can accurately extract the corral radius in parameter regimes commonly encountered in membrane diffusion studies. The domain of validity was also discussed.  

Two popular data generating processes were considered (reflected Brownian motion and the OU process).  It was demonstrated that accurate results can be obtained even if the stochastic model assumed was not consistent with the  data generating process providing some robustness assurance.  In the confinement regime and sample sizes considered, there was not adequate evidence to distinguish reflected Brownian motion from an OU process.  
The estimated corral radius was reliably extracted using an OU model regardless of the underlying stochastic dynamics and measurements producing the observational data.
  Numerous statistically motivated reasons for favoring the OU model to the reflected Brownian motion model were discussed.
Potential problems that can be encountered when the inferred corral radius is too large to reliably infer
from the data available were also discussed. The rich likelihood structure afforded by wrapping SDE plus noise models around experimental data was exploited throughout.  
The likelihood formulation circumvents 
the need for selecting  \emph{ad hoc} sampling parameters such as a ``time-lag" cut-off (this is a common problem  in MSD-based analysis \cite{Michalet2012}; MSD-based analyses are still quite popular in the SPT community).

The ability to accurately extract kinetic parameters  and correct for biases induced by small time series sample sizes (while also accounting for measurement and thermal noise in a statistically rigorous fashion \cite{SPAfilter, Berglund2010}) shows great promise studies where the underlying molecule experiences random forces whose distribution  changes in both time and space due to complex interactions in a highly heterogeneous environment.
For example, if one can both reliably determine when molecules leaves a ``picket fence" \cite{Kusumi2005} in the plasma membrane via
 change point detection algorithms \cite{Poor2008} and can track trajectories with high temporal resolution (perhaps at the cost of spatial accuracy), one can utilize the tools presented here to accurately map out both the diffusion coefficient  and the corral radii explored by molecules in the plasma membrane or in the cytoplasm \cite{Thompson2010}.  This presents an attractive physically interpretable modeling alternative to  sub-diffusion or continuous time random walk type models, but such a study is left to future work.  The method introduced was shown to be useful in parameter regimes commonly encountered in fluorescence-based SPT experimental studies, but the approach is general and can be used to probe other length and time scales. 

 
\section{Acknowledgements}
The author would like to thank Randy Paffenroth (Numerica Corp.) for comments on an earlier draft. \\

\section{Appendix} 
\subsection{MSD of the Stationary ($\kappa > 0$) OU Process}
The MSD associated with $\delta$ time units between observations is defined by $\frac{1}{N} \langle \sum \limits_{t=1}^N (x_{t+\delta }-x_{t})^2 \rangle$; in the previous expression  $\langle \cdot \rangle$ denotes ensemble averaging \cite{Park2010,Magdziarz2010}.  Let $\mathcal{M}(\delta)$ denote  the MSD multiplied by $N$ at a given lag $\delta$;  then plugging in the solution to the mean zero stationary OU process  (variance $=\frac{\sigma^2}{2\kappa}$ \cite{risken}) and exploiting other standard properties of SDEs driven by Brownian motion \cite{kp} yields:

\begin{align} 
\nonumber \mathcal{M}(\delta)= & \langle \sum\limits_{t=1}^N   \big(x_{t}e^{-\kappa \delta} + \sigma \int \limits_0^\delta e^{-\kappa(\delta-s)}dW_s\big) -x_{t})^2 \rangle \\
\nonumber = &\langle \sum   x_{t}^2e^{-2\kappa \delta} + x_{t}^2 + \frac{\sigma^2}{2\kappa}(1-e^{-2\kappa\delta})  -2x_t^2e^{-\kappa \delta} \rangle \\
\nonumber = &\langle \sum  x_{t}^2\big(1+e^{-2\kappa \delta}-2e^{-\kappa \delta} \big) + \frac{\sigma^2}{2\kappa}(1-e^{-2\kappa\delta}) \rangle \\
\nonumber = & \sum  \langle x_{t}^2 \rangle \big(1+e^{-2\kappa \delta}-2e^{-\kappa \delta}\big) + \frac{\sigma^2}{2\kappa}(1-e^{-2\kappa\delta})  \\
\nonumber = & \sum   \frac{\sigma^2}{2\kappa} \big(1+e^{-2\kappa \delta}-2e^{-\kappa \delta}\big) + \frac{\sigma^2}{2\kappa}(1-e^{-2\kappa\delta}) \\
  =& \sum\limits_{t=1}^N   \frac{\sigma^2}{\kappa} \big(1-e^{-\kappa \delta}\big) 
\label{eq:MSDOU}
\end{align}

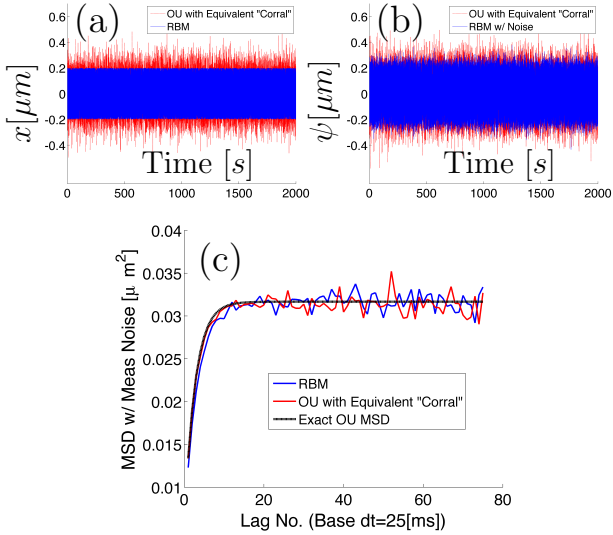
\begin{figure}[htb]
\center
\centering
\begin{minipage}[b]{.95\linewidth}
\def\pw{.475}
\begin{overpic}[width=\pw\textwidth]{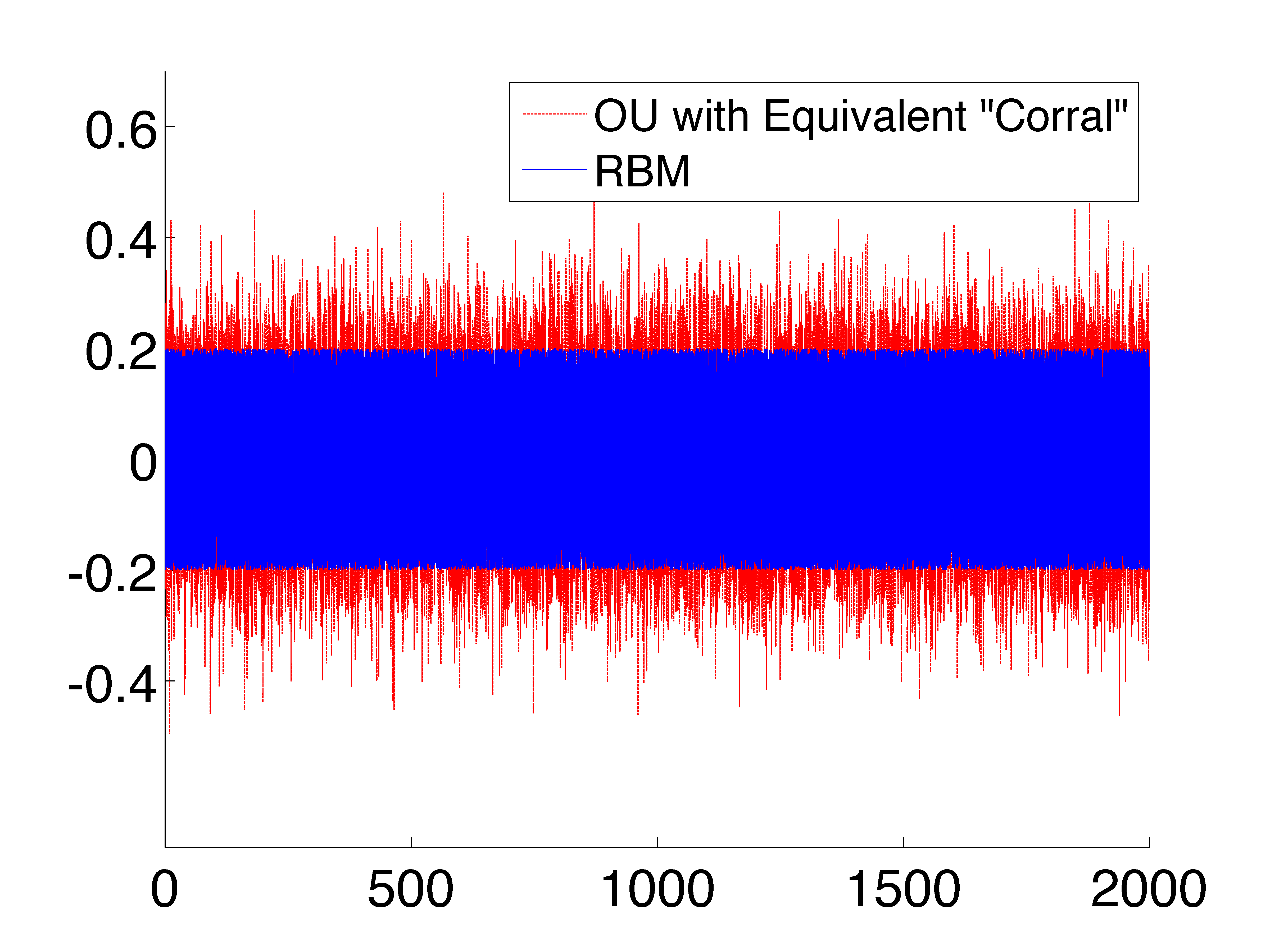}
\put(15,60){\Large \color{black}(a)}
\put(-8,22){\rotatebox{90} {\makebox[1.\width]{\large $x [\mu m]$} \hfill }}
\put(38,12){\rotatebox{0} {\makebox[1.\width]{\large Time $[s]$} \hfill }}
\end{overpic}
\begin{overpic}[width=\pw\textwidth]{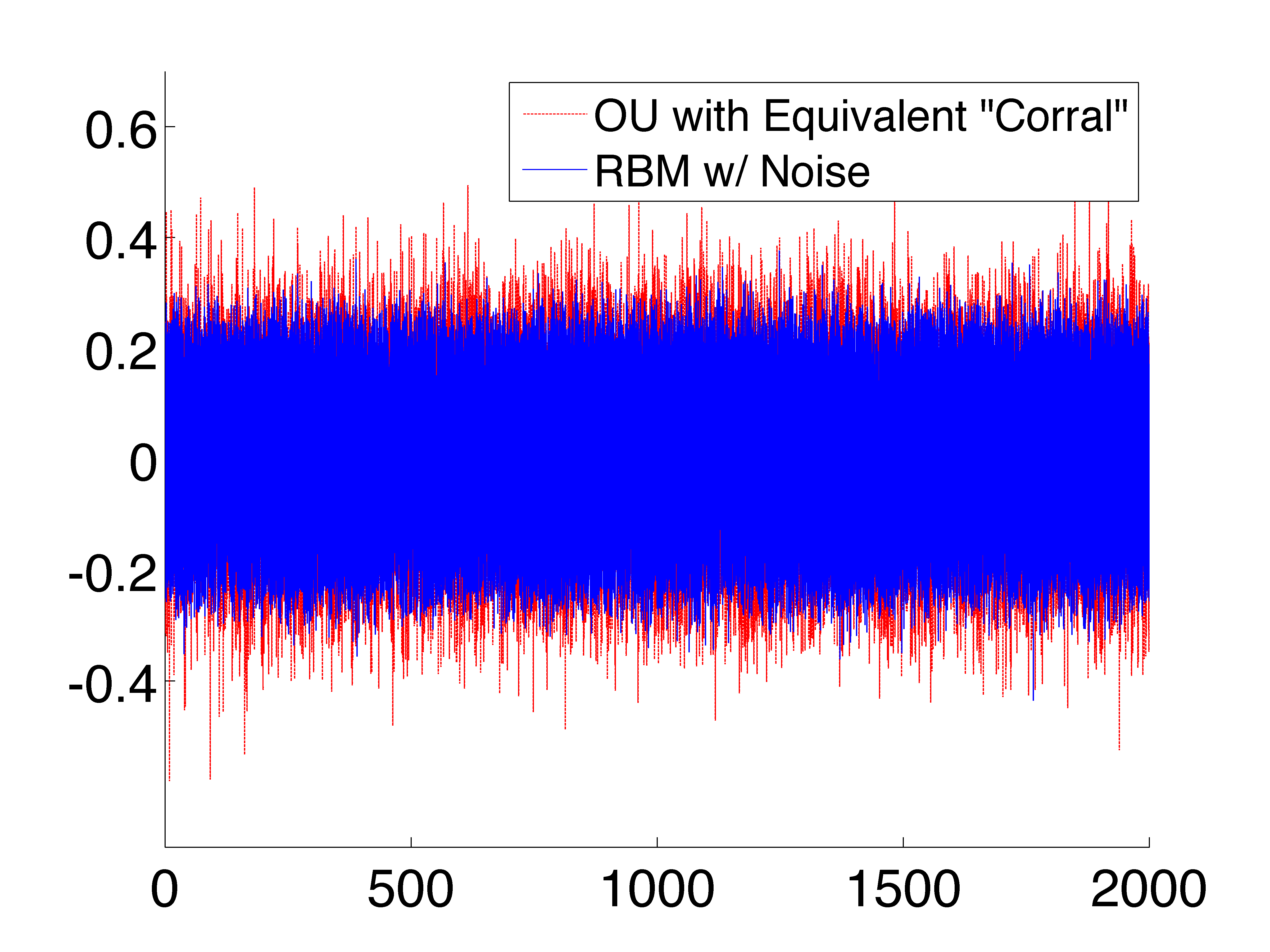}
\put(15,60){\Large \color{black}(b)}
\put(-8,22){\rotatebox{90} {\makebox[1.\width]{\large $\psi [\mu m]$} \hfill }}
\put(38,12){\rotatebox{0} {\makebox[1.\width]{\large Time $[s]$} \hfill }}
\end{overpic}

  \end{minipage}
\begin{minipage}[b]{.95\linewidth}
\def\pw{.7}
  \begin{overpic}[width=\pw\textwidth]{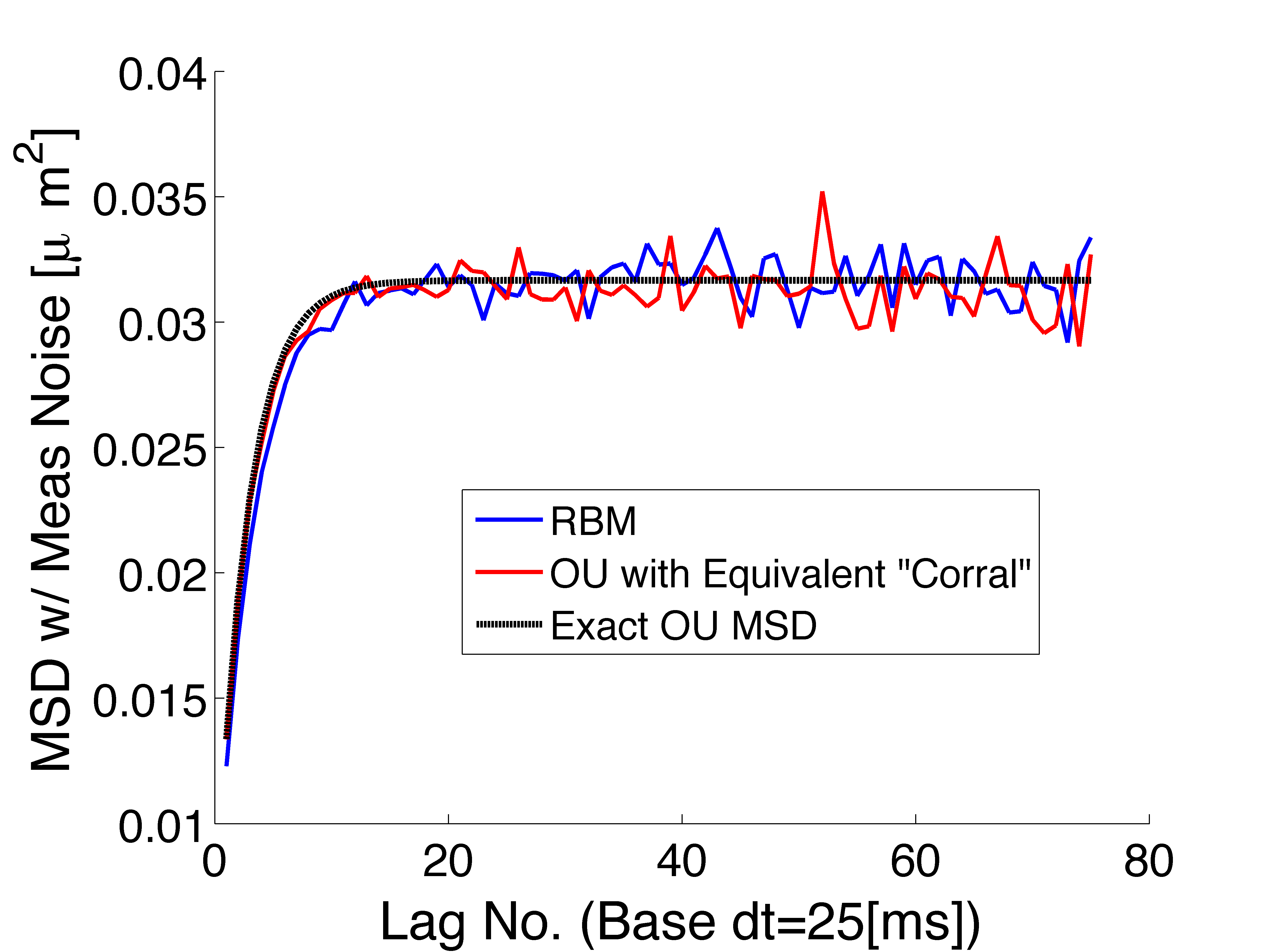}
\put(20,60){\Large \color{black}(c)}
\end{overpic}
  \end{minipage}

\centerline{\footnotesize }
\caption{
\footnotesize 
(Color online) 
Long time sample (hence large $N$) trajectory for OU and RBM process without (a) and with (b) measurement noise.  Panel (c) shows that for large $N$, the empirical MSD curve matches the theoretical (i.e., infinite sample limit) MSD limit.  Here $L=400nm, D=0.2\mu m^2/2, R=50nm$, and $\Delta t=25 ms$.
 }
\label{fig:A1}
\end{figure}

\begin{figure}[htb]
\center
\centering

\begin{minipage}[b]{.95\linewidth}
\def\pw{.475}
\begin{overpic}[width=\pw\textwidth]{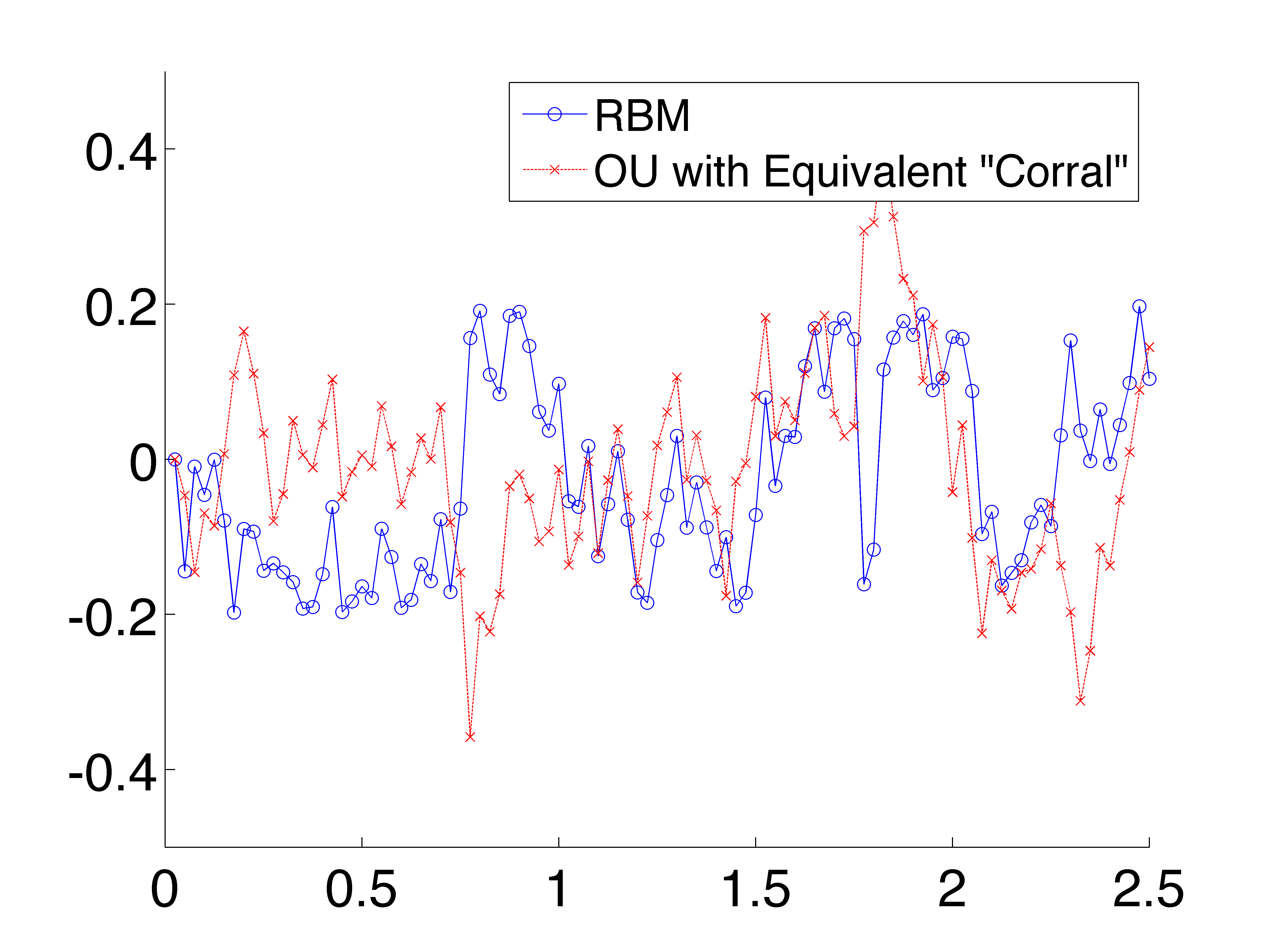}
\put(15,60){\Large \color{black}(a)}
\put(-8,22){\rotatebox{90} {\makebox[1.\width]{\large $x [\mu m]$} \hfill }}
\put(38,12){\rotatebox{0} {\makebox[1.\width]{\large Time $[s]$} \hfill }}
\end{overpic}
\begin{overpic}[width=\pw\textwidth]{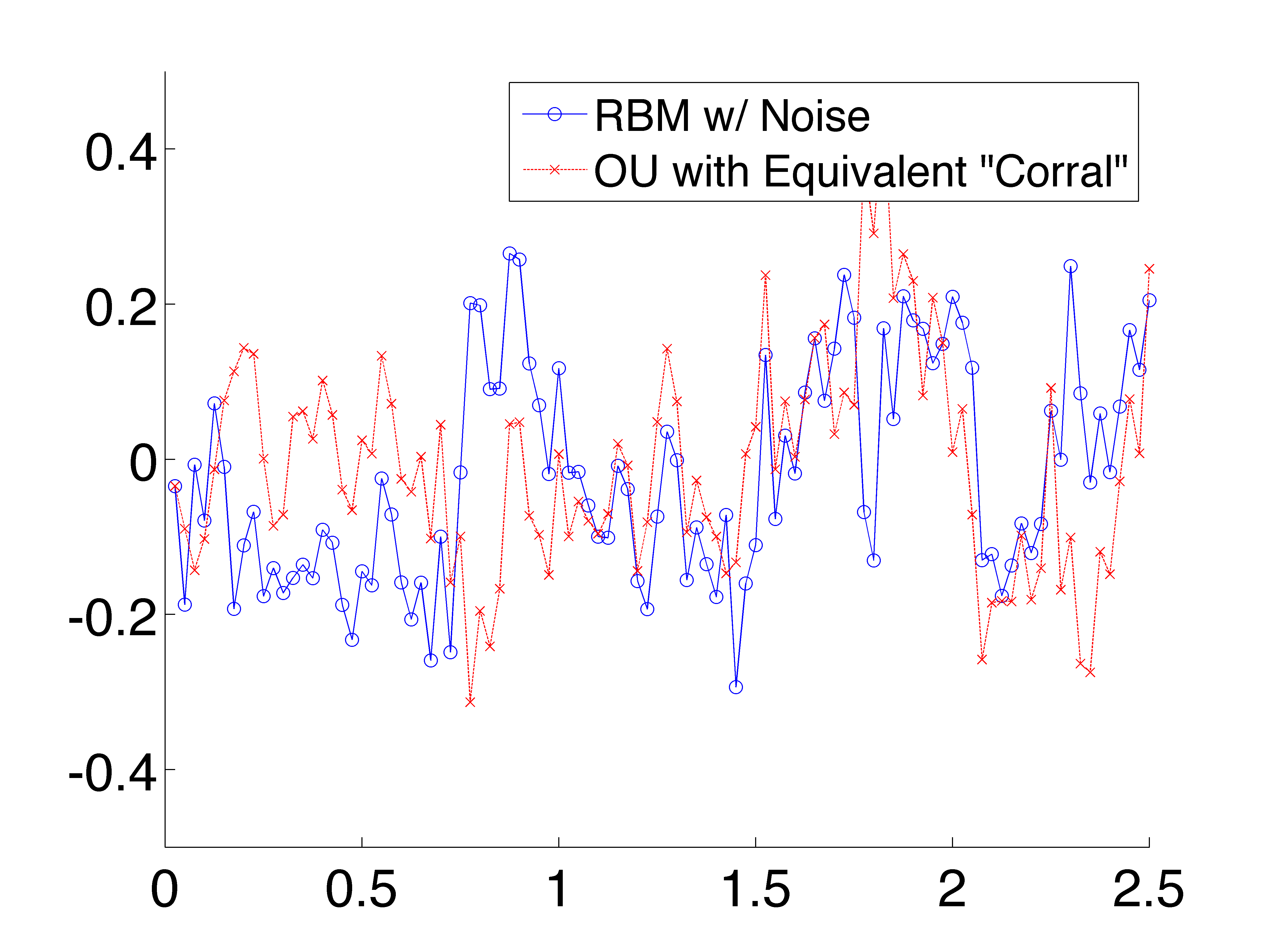}
\put(15,60){\Large \color{black}(b)}
\put(-8,22){\rotatebox{90} {\makebox[1.\width]{\large $\psi [\mu m]$} \hfill }}
\put(38,12){\rotatebox{0} {\makebox[1.\width]{\large Time $[s]$} \hfill }}
\end{overpic}
  \end{minipage}
\begin{minipage}[b]{.95\linewidth}
\def\pw{.7}
  \begin{overpic}[width=\pw\textwidth]{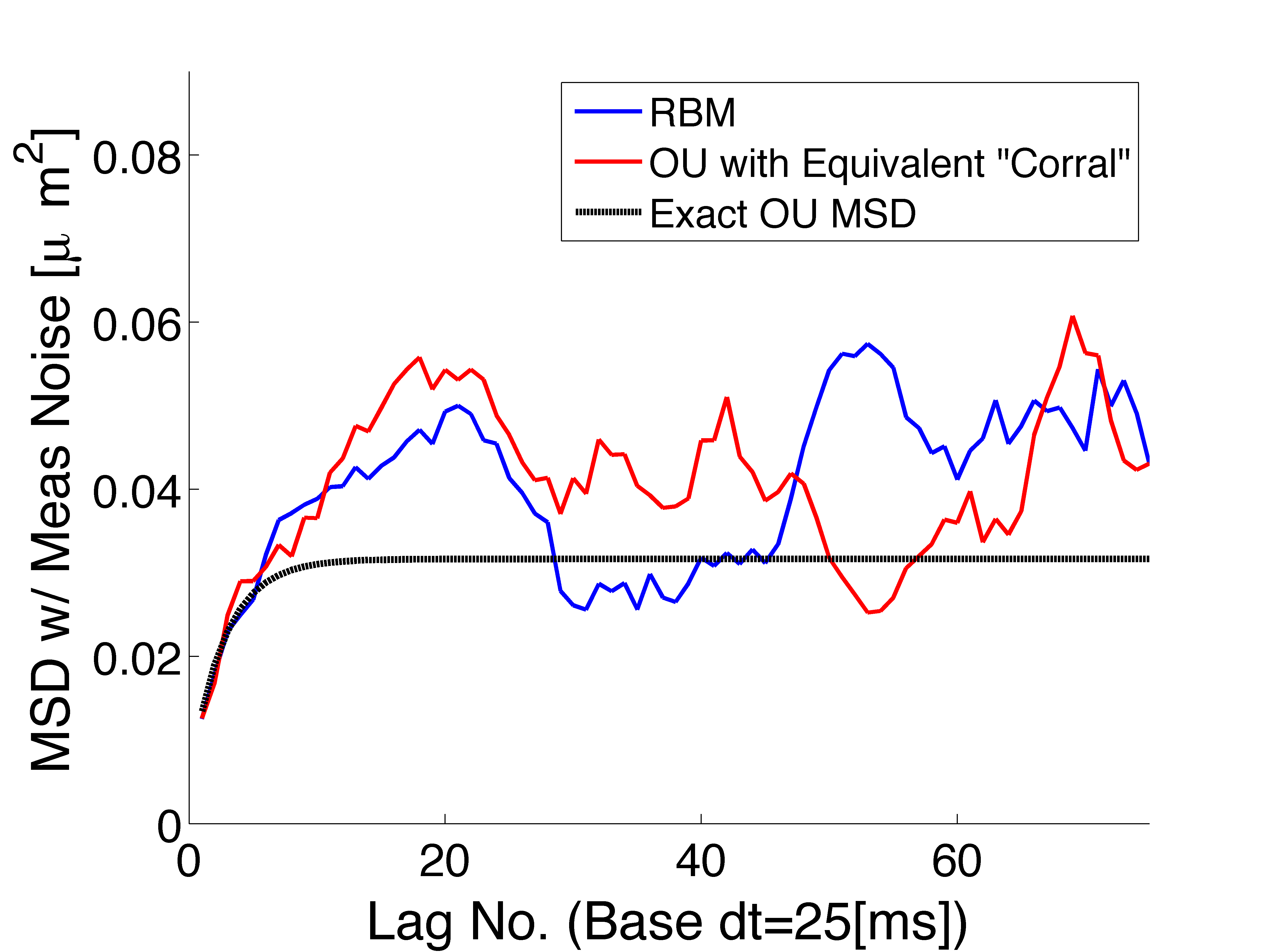}
\put(20,60){\Large \color{black}(c)}
\end{overpic}
  \end{minipage}

\centerline{\footnotesize }
\caption{
\footnotesize 
(Color online) 
Same as Fig. \ref{fig:A1}, except the sample size was reduced to $N=100$ observations.  The small sample size complicates reliably using an MSD-based analysis. With short track lengths (small $N$), well-known issues associated with selecting the lag truncation to use in computations, statistical dependence commonly introduced when computing MSDs, etc. \cite{Michalet2012} are even more pronounced.  The likelihood-based bias correction scheme introduced is able to reliably extract system parameters even with these small samples sizes.
 }
\label{fig:A2}
\end{figure}
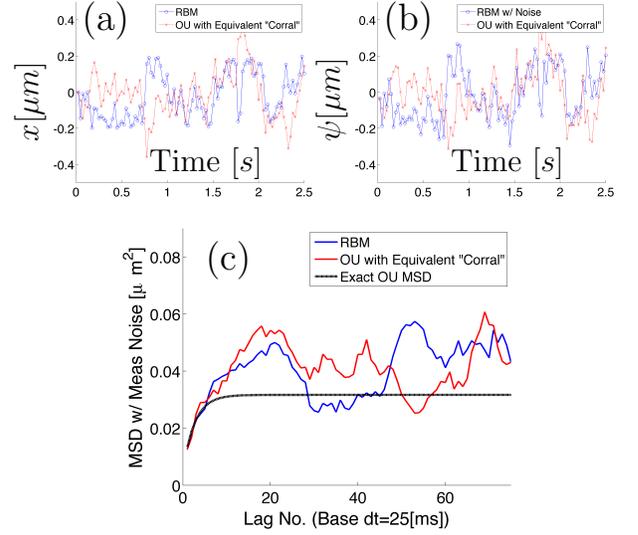

To account for i.i.d. Gaussian measurement noise (i.e., one carries out an MSD on $\psi$) in the above expression, simply add $2\times R\ \times N$ to the MSD expression above \cite{TSRV}.

\subsection{Representative Trajectories and MSDs}
In this section, the reflected Brownian motion and the corresponding OU process (found using Eqn. \ref{eq:bias}) are plotted with and without measurement noise.  The MSDs of the measurement noise free and measurement noise case are shown for both large and small sample sizes.

\subsection{MLE of the Innovation Sequence}
\label{A:innov}
In the main text, mappings between the OU parameters and those of the classic Kalman filter \cite{hamilton} were presented.
Here the equations defining the innovation MLE and the associated likelihood \cite{hamilton,SPAdsDNA} relevant to the scenario studied are presented (the reader is referred to Ref. \cite{hamilton} for full details).  Note that the ``observation matrix'' $H$ is the identity matrix and that
$\hat{\psi}_{i|i-1}\equiv H\hat{x}_{i|i-1}$ (= $\hat{x}_{i|i-1}$ in the case considered). 

\begin{flalign}
\label{eq:innov}
\\
 \nonumber (\hat{R},\hat{F},\hat{Q}) =  \mathrm{argmax} \ \mathcal{L}(R,F,Q) \equiv p(\psi_1,\psi_2,\ldots,\psi_N; R,F,Q) \hfill \\
\nonumber p(\psi_1,\psi_2,\ldots,\psi_N; R,F,Q)= \\
\nonumber\prod\limits_{i=1}^{N}\frac{1}{\sqrt{2\pi S_i}}\exp \big( \frac{-(\psi_i-F\hat{\psi}_{i-1|i-1})^2}{2 S_i} \big) \\
\nonumber \hat{\psi}_{i|i}=\hat{\psi}_{i|i-1}+K_i(\psi_i-\hat{\psi}_{i|i-1}) \\
\nonumber K_i=\frac{P_{i|i-1}}{P_{i|i-1}+R} \\
\nonumber S_i=P_{i|i-1}+R \\
\nonumber P_{i|i-1}=FP_{i-1|i-1}F + Q \\
 \nonumber P_{i|i}=P_{i|i-1}-\frac{P_{i|i-1}^2}{P_{i|i-1}+R}
\end{flalign}

For the stationary OU process, the recursion above (processing the observation sequence) was started using $\hat{x}_{1|0}=0$ and $P_{1|0}=\frac{\sigma^2}{2\kappa}$.   The Nelder-Mead algorithm was used to find the parameter optimizing Eqn. \ref{eq:innov}.  Goodness-of-fit testing \cite{hong,SPAgof} was used to both check the consistency of model assumptions against data and to ensure that a local minimum was not encountered in the optimization. 

\bibliographystyle{nature}
\bibliography{StatPapers,biomedtracking,tracking,running,ion_channels,Primary_Cilium}

\end{document}